\documentstyle[11pt,Moriond]{article}
\begin{document}
\vspace*{4cm}
\title{FIELD-DEPENDENT  BRS TRANFORMATIONS 
AND CORRECT PRESCRIPTION FOR ${1\over{(\eta\cdot k)^p}}$-TYPE 
SINGULARITIES IN AXIAL GAUGES}

\author{A. Misra}

\address{Institute of Physics, Sachivalaya Marg, Bhubaneswar 751005, India}

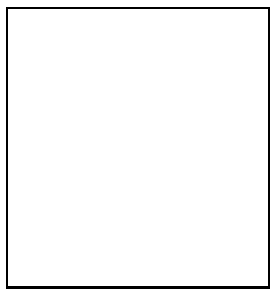
\maketitle\abstracts{
The axial-gauge boson propagator contains $1/(\eta\cdot k)^p$-type
singularities. These singualarities have generally been treated by   
inventing prescriptions for them. We propose an alternative procedere
for treating these singularities in the path-integral formalism using
the known way of treating the $1/k^{2n}$-type singularities
in Lorentz-type gauges. For this purpose
we use a finite field-dependent
BRS transformation that inerpolates between the Lorentz and
axial-type gauges. We arrive at the $\epsilon$-dependent
tree propagator in axial-type gauges.}

Calculations in nonabelaian gauge theories require a choice of 
gauge.  There are many families of
gauges that have been used in practical calculations, e.g., 
Lorentz-type gauges, axial gauges.
{\it It  becomes an 
important question as to how the calculations in various (families of)
gauge choices are related to each other.}
Gauge-independence in a limited framework,
has been proven in early days. 
Such proofs utililize the {\it infinitesimal} gauge transformations reponsible
for gauge-parameter change. Ways of connecting Green's
functions in   family of gauges (and establishing expliticly gauge-independence)
has not been done until recently. Discrepancies have
been reported in anomalous dimension calculations in the Lorentz-type and
axial-type gauges.
Thus, it becomes important to obtain a procedure to connect the Green functions 
in different families of gauges.

Unlike gauge transformations, ``infinitesimal" and ``finite"
BRS transformations have the same form. This makes them suitable
for our purpose.  The FP effective actions in the Lorentz and Axial gauges 
are invariant under the BRS transformations:
\begin{equation}
\label{eq:BRS1}
\delta\phi_i(x)=\delta_{\rm BRS}\phi_i(x)\delta\Lambda
\end{equation}
where $\delta_{\rm BRS}\phi_i(x)$ is given in \cite{jm}$^)$, and
the gauge and (anti)ghost fields are generically denoted by
$\phi_i$.

As observed by Joglekar and Mandal \cite{jm}$^)$,
$\delta\Lambda$ {\it need not be infinitesimal
 nor need it be field-independent as long as it  does not  depend on
$x$ explicitly for (\ref{eq:BRS1}) to
be a symmetry of FPEA}
In fact, the following finite field-dependent BRS (FFBRS)
transformations were introduced:
\begin{equation}
\label{eq:BRS3}
\phi^\prime_i(x)=\phi_i(x)+\delta_{\rm BRS}\phi_i(x)\Theta[\phi],
\end{equation}
where $\Theta[\phi]$  
is an $x$-independent functional of $A,\ c,\ {\bar c}$
and these were also the symmetry of the FPEA.  

The FPEA is invariant under (\ref{eq:BRS1}), but the
functional measure is not invariant under the
(nonlocal) transformations (\ref{eq:BRS1}). The Jacobian for the
FFBRS transformations can be expressed (in special
cases effectively as
$exp(iS_1)$ and this $S_1$ explains the difference between the two
effective actions \cite{jm}$^{),\ }$\cite{bj}$^)$. 

Such FFBRS transformations were constructed  
by integration of an infinitesimal field-dependent BRS
(IFBRS) transformation:
\begin{equation}
\label{eq:BRS4}
{d\phi_i(x,\kappa)\over{d\kappa}}
=\delta_{\rm
BRS}[\phi(x,\kappa)]\Theta^\prime[\phi(x,\kappa)].
\end{equation}
The integration of (\ref{eq:BRS4}) from $\kappa=0$ to 1, 
leads to the FFBRS transformation of (\ref{eq:BRS3}) 
with $\phi(\kappa=1)\equiv \phi^\prime$ and $\phi(\kappa=0)=\phi$.
The relationship between $\Theta$ and $\Theta^\prime$ is given in ref 1.

The result for the FFBRS Transformation for Correlating Green's functions in 
Lorentz-type Gauges to Axial-type  Gaugees (that differs from the
FFBRS of refs. 1 and 2) reads \cite{jajmp}$^)$:
\begin{eqnarray}
\label{eq:result3}
& & \langle O[\phi]
\rangle_A\equiv\int{\cal D}\phi^\prime O[\phi^\prime]
e^{iS_{\rm eff}^A[\phi^\prime]}\nonumber\\
& & =\int{\cal D}\phi{\cal O}[\phi]e^{iS^L_{\rm eff}[\phi]}+
\int{\cal D}\phi\sum_i\delta\phi_i[\phi]
{\delta{\cal O}\over{\delta\phi_i}} e^{iS^L_{\rm eff}},
\end{eqnarray}
where the summation over $i$ runs over fields $A, c, {\bar  c}$ and
\begin{eqnarray}
\label{eq:result1}
& & \phi^\prime=\phi+\biggl(\tilde\delta_1[\phi]\Theta_1[\phi]
+\tilde\delta_2[\phi]\Theta_2[\phi]\biggr)\Theta^\prime[\phi]
\nonumber\\
& & \equiv\phi+\delta\phi[\phi]
\end{eqnarray}
is an FFBRS  with
\begin{equation}
\label{eq:Theta12def}
\Theta_{1,2}[\phi]\equiv\int_0^1 d\kappa (1,\kappa)exp\biggl(\kappa
f_1[\phi]
+{\kappa^2\over 2}f_2[\phi]\biggr);
\end{equation}
\begin{equation}
\label{eq:f12def1}
f[\tilde\phi,\kappa]\equiv
f_1[\tilde\phi]+\kappa f_2[\tilde\phi];
\end{equation}
\begin{eqnarray}  
\label{eq:f12def2}
& & f_1[\phi]\equiv i\int d^4x\biggl[{\partial\cdot A^\alpha\over\lambda}
(\partial\cdot A^\alpha-\eta\cdot A^\alpha)+{\bar c}
(\partial\cdot{\rm D}-\eta\cdot{\rm D})c\biggr]
\nonumber\\
& & f_2[\phi]\equiv -{i\over\lambda}
\int d^4x (\partial\cdot A^\alpha-\eta\cdot A^\alpha)^2,
\end{eqnarray}
and
\begin{equation}
\label{Thprimedef}
\Theta^\prime\equiv i\int d^4x\ {\bar c}^\alpha
(\partial\cdot A^\alpha-\eta\cdot A^\alpha).
\end{equation}  
An alternate and more effective expression can be given:
\begin{equation}
\label{eq:AtoL2}
\langle {\cal O}\rangle_A=\langle{\cal O}\rangle_L+\int_0^1 d\kappa\int
D\phi
\sum_i\biggl(\tilde\delta_{1,i}[\phi]+\kappa\tilde\delta_{2,i}[\phi]
\biggr)\Theta^\prime[\phi]
{\delta{\cal O}\over{\delta\phi_i}} e^{iS^M_{\rm eff}},
\end{equation}
where $\tilde\delta_{1,i}$ and $\tilde\delta_{2,i}$ are defined in
via
\begin{equation}  
\label{eq:tildedelsdef}
\tilde\delta_{\rm BRS}\phi_i\equiv
\biggl(\tilde\delta_{1,i}+\kappa\tilde\delta_{2,i}\biggr)\delta\Lambda
\end{equation}
where $\tilde\delta_{\rm BRS}\phi_i$ are the BRS variations for the mixed
gauge function
[$\partial\cdot A(1-\kappa)+\kappa\eta\cdot A$].

The basic idea is to use (\ref{eq:AtoL2}) to relate the axial and
Lorentz gauge propagators. The only shortcoming of the above relation
is that it does not include the
$i[-\epsilon A^2/2+\epsilon {\bar c}c]$ terms in the Lorentz gauge
effective action.  The modification  of (\ref{eq:AtoL2}) is:
\begin{eqnarray}
\label{eq:presp15}
& & \langle O\rangle_A = \langle O\rangle_L\nonumber\\
& & +\int_0^1 d\kappa\int{\cal D}\phi e^{i[S^M_{\rm eff}[\phi,\kappa]
-i\epsilon(A^2/2-{\bar c}c)]}\biggl[\biggl(\delta_1[\phi]
+\kappa\delta_2[\phi]\biggr)\Theta^\prime[\phi]
{\delta O\over{\delta\phi}}\biggr].
\end{eqnarray}
Thus, in {\it this form}, the only effect on the second term is
to modify $S_{\rm eff}^M$ by
$-i\epsilon\int (A^2/2-{\bar c}c)d^4x$ inside $\kappa$-integration.

We now employ the result (\ref{eq:presp15}) for the propagators. We
set $O[\phi]= A^\alpha_\mu(x)A^\beta_\nu(y)$. The equation
(\ref{eq:presp15}) then reads:
\begin{eqnarray}
\label{eq:exres}
& & iG^{0A\ \alpha\beta}_{\mu\nu}(x-y)=iG^{0L\ \alpha\beta}_{\mu\nu}(x-y)
+i\int_0^1 d\kappa\int{\cal D}\phi e^{i[S_{\rm eff}^M[\phi,\kappa]-i
\epsilon\int(A^2/2-{\bar c}c)d^4x]}\nonumber\\
& & \times
\biggl(({\rm D}_\mu c)^\alpha(x)A^\beta_\nu(y)
+A^\alpha_\mu(x)
({\rm D}_\nu c)^\beta(y)\biggr)
\int d^4z{\bar c}^\gamma(z)(\partial\cdot A^\gamma-\eta\cdot A^\gamma)(z).
\nonumber
\end{eqnarray}
This leads to, for zero loop case \cite{jampla1}$^)$,
\begin{equation}
\label{eq:freslt1}
\tilde G^{0A}_{\mu\nu}=\tilde G^{0L}_{\mu\nu}
+\biggl[\biggl(k_\mu k_\nu\Sigma_1+\eta_\mu k_\nu\Sigma_2\biggr)ln\Sigma_3
+(k\to -k;\mu\leftrightarrow\nu)\biggr]
\end{equation}
where
\begin{eqnarray}
\label{eq:freslt2}
& & \Sigma_1\equiv {-(k^2-i\eta\cdot k)\biggl({\eta\cdot k+i\eta^2
\over{k^2-i\eta\cdot k}}+i\lambda
-{(1-\lambda)\eta\cdot k\over{k^2+i\epsilon}}\biggr)
\over{\epsilon\Sigma}}\nonumber\\
& & \Sigma_2\equiv {-(k^2-i\eta\cdot k)\biggl(
-\biggl[{k^2+i\eta\cdot k\over{k^2-i\eta\cdot k}}
\biggr]+1-{i\epsilon(1-\lambda)\over{k^2+i\epsilon}}\biggr)
\over{\epsilon\Sigma}}\nonumber\\
\nonumber\\
& & \Sigma_3\equiv{-i(\eta\cdot k+\epsilon)(k^2+i\epsilon\lambda)
\over{(k^2+i\epsilon)\biggl(-i\epsilon\lambda-\sqrt{k^4-(k^2+i\epsilon\lambda)
\biggl[k^2+{(\eta\cdot
k)^2+i\epsilon\eta^2\over{k^2+i\epsilon}}\biggr]}\biggr)}},\nonumber\\
& & {\rm and}\nonumber\\
& & \Sigma\equiv\biggl[(1-\lambda)[(\eta\cdot k)^2+2ik^2\eta\cdot k]
+i\epsilon
k^2(1-2\lambda)+\lambda(k^2+i\epsilon)^2+\eta^2(k^2+i\epsilon)\biggr].
\nonumber\\
& &
\end{eqnarray}

The $k^0$ integration over this
propagator can be replaced by a $k^0$-integration over (most of) the real
axis combined over semicircle in the LHP of radius $>>\sqrt{\epsilon}$
(where the complication due to presence of $\epsilon $ can
be dropped and the usual simple form can be used)
and an  additional effective term of much simpler form that rounds up
effectively the  complex structure near $\eta\cdot k=0$ 
\cite{jaijmpa}$^)$.
For $\eta^2\neq0$, and $k^2\neq0$, the latter reads
\begin{equation}
\label{eq:fresult3}
\delta\biggl(k^0-{1\over2}\sqrt{{\epsilon\eta^2\over i}}
-\vec\eta\cdot\vec k\biggr)
\biggl[k_\mu k_\nu D_1+\eta_\mu k_\nu D_2+\eta_\nu k_\mu D_3\biggr]
+\mu\leftrightarrow\nu;k\rightarrow-k
\end{equation}
\begin{eqnarray}
\label{eq:fresult4}
& & D_1\equiv -{\pi\eta^2\over{\cal K}_1}
i\sqrt{{i\eta^2\over\epsilon}}
+{i\pi(\eta^2)^2{\cal K}_2\over{2{\cal K}_1^2}}
;\nonumber\\
& & D_2\equiv {i\pi\eta^2\over{2{\cal K}_1}};\
D_3\equiv -{i\pi\eta^2\over{2{\cal K}_1}},
\end{eqnarray}
where
\begin{eqnarray}
\label{eq:simpeq14}
& & {\cal K}_1\equiv\biggl((\vec\eta\cdot\vec k)^2-\vec k^2\biggr)
(\eta^2+i\epsilon);
\nonumber\\
& & {\cal K}_2\equiv 2i\biggl((\vec\eta\cdot\vec k)^2-\vec k^2\biggr)
+2\vec\eta\cdot\vec k(\eta^2+i\epsilon).
\end{eqnarray}
Note {\it that if we define the LCG as the $\eta^2\rightarrow0$
limit, then this additional term (\ref{eq:fresult3}) vanishes}. Thus,
we obtain a simple result of the LCG.

For $|\eta\cdot k|>>\epsilon$,
(\ref{eq:freslt1}) gives the expected result:
\begin{equation}
\label{eq:eps=0}
\tilde G^{0 A}_{\mu\nu}-\tilde G^{0 L}_{\mu\nu}=
-{1\over k^2}k_\mu k_\nu
\biggl({(\lambda k^2+\eta^2)\over{(\eta\cdot k)^2}}+{(1-\lambda)\over k^2}
\biggr) + {k_{[\mu}\eta_{\nu]_+}\over{k^2\eta\cdot k}}.
\end{equation}

As the FFBRS transformations discussed also  preserve the vacuum
expectation values of gauge-invariant observables,
it follows that our treatment is such that by its very construction,
the Wilson loop $W[L]$ has the same value in the Lorentz and axial-type
gauges to {\it all} orders.
We have further given the proof of this
statement to O$(g^4)$  using the earlier work by   
Cheng and Tsai. Our proof holds for any arbitrary  loop for
$\eta^2<0$ and for a subclass of loops for $\eta^2\geq0$
\cite{jampla2}$^)$.

We also note that the O$(g^2)$ thermal Wilson loop $W_R$ \cite{jampla2}$^)$, 
depends only on $D_{00}(k^0=0,\vec k)$. We find
that for the propagator in (\ref{eq:freslt1}),
$D_{00}(k^0=0,\vec k)={g_{00}\over{k^2+i\epsilon}}$ which
is the same as $D_{00}(k^0=0,\vec k)$  for Lorentz
gauges and as such $W_R$ has the same value as in Lorentz-type gauges.

In conclusion, we addressed the
problem of relating arbitrary Green's functions in two sets of uncorrelated
gauges, e.g..  the axial and the Lorentz-type
gauges (the example considered). 
We showed that this involved an
FFBRS, obtained by intregration of an IFBRS.
We found that the final result could be put in a neat form
(\ref{eq:result3}) or (\ref{eq:AtoL2}). 
Using our result, we have $derived$ the correct prescription 
for Axial gauge propagator.
Even though, the propagator in axial gauge, naively calculated, has
spurious singularities.
the correct treatment of these singularities
is obtained by relating this propagator to the
corresponding Lorentz gauge treatment. This
was  done by using the FFBRS discussed in this talk.
The propagator of (\ref{eq:freslt1}) gives, however complex, the
actual correct treatment of these singularities.
While for $|\eta\cdot k|>>\epsilon$, it
gives the usual propagator, the actual analytic nature of the propagator,
in the vicinity of the origin is much more
complicated than indicated by various
prescriptions suggested earlier.

\section*{Acknowledgement}
I wish to thank my collaborator Prof.S.D.Joglekar (IIT Kanpur, India)
for introducing me to FFBRS transformations and
sharing his insights with me. I also wish to thank the Moriond
committee for inviting me to the conference.

\end{document}